# Using the gameplay and user data to predict and identify causes of cybersickness manifestation in virtual reality games

1st Thiago Porcino
*Institute of Computing*
*Universidade Federal Fluminense*
*Firjan - ISI SVP*
Rio de Janeiro, Brazil
tmalheiros@id.uff.br

2nd Erick O. Rodrigues
*Academic Department of Informatics*
*Universidade Tecnologica Federal do Paraná*
Pato Branco, Parana
erodrigues@utfpr.edu.br

3rd Alexandre Silva
*Departament of Academic Informatics*
*Instituto Federal do Triangulo Mineiro*
Uberaba, Brazil
alexandre@iftm.edu.br

4th Esteban Clua
*Institute of Computing*
*Universidade Federal Fluminense*
Niterói, Brazil
esteban@ic.uff.br

5th Daniela Trevisan
*Institute of Computing*
*Universidade Federal Fluminense*
Niterói, Brazil
daniela@ic.uff.br

***Abstract*—Virtual reality (VR) is an imminent trend in games, education, entertainment, military, and health applications, as the use of head-mounted displays is accessible to everyone. While VR provides immersive experiences, it still does not offer an entirely perfect situation, mainly due to cybersickness (CS) issues. In this work, we propose a novel approach for predicting upcoming CS symptoms. Our solution is able to suggest whether the user of VR is entering into an illness situation. We adopted random forest classifiers and validated our solution using 16 different machine-learning techniques, which presented the best results. For training purposes, we built our own dataset through a CS profile questionnaire that we also propose in the present work. The questionnaire is focused on registering and identifying the user's susceptibility to CS, considering their historical conditions and also their response to the immersive environment developed by us. In this method, 86 individuals are selected and the developed questionnaire was put to them on different days, and the answers are compiled as dataset. Our proposal also identifying attributes responsible (causes and individual's parameters) for the observed stressful and uncomfortable situations.**

## I. Introduction

We are currently experiencing the growth and consolidation of a new communication platform. Virtual reality (VR) delivers immersive 3D graphics in entertainment applications, serious games or applied games, and training applications in health, technological, military, or scientific domains.

Meanwhile, most users who experience (use) head-mounted displays activities feel one or more symptoms of sickness, especially when the user is subjected to it (using it) for a long period of time [24]. According to Ramsey et al. [36], on average, eighty percent of participants who experienced VR with head mounted displays (HMDs) felt discomfort after the first 10 min of exposure. Therefore, more and extensive VR activities tend to cause stronger discomfort levels. However, the level of discomfort will vary according to the individual profiles.

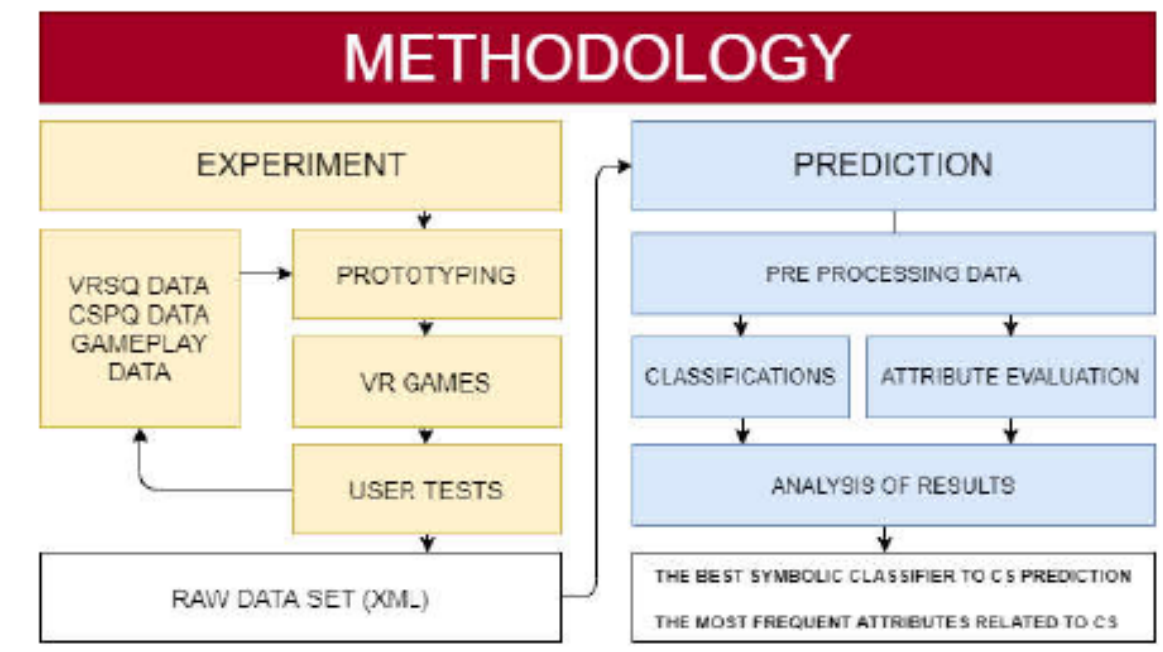

Fig. 1: The methodology proposed in this work.

According to Hua [15], minimizing sickness in virtual and augmented reality applications is an unresolved challenge. Discomfort resulting from VR can arise from three different leading causes: motion sickness (MS), visually induced motion sickness (VIMS), and cybersickness (CS) or simulator sickness (SS).

Motion sickness is manifested by the divergence of information emitted by the human sensory system. This happens when conflicts between the sensory organs when they try to define the orientation and position in space. MS Sickness is defined as the manifestation of discomfort during a forced visual movement but without the movement of the body, such as during airplane flights, boat trips, or even with land vehicles [3], [46]. Here our body's sensory organs send mixed messages to the brain, causing dizziness, lightheadedness, or nausea. This uncomfortable experience also occurs in virtual environments and it is called VIMS.

VIMS symptoms are similar to MS. However, the difference lays on the fact that there is no physical movement in VIMS or they are extremely limited [19]. Several studies categorized VIMS as MS symptoms when situations of visual stimulation manifest irrespective of physical movements. Due to that, different studies in different VIMS contexts were renamed to match the set of symptoms according to the environment they manifested.

Merhi et al. [30] define the occurrence of VIMS during experiments with video games as game sickness. Brooks et al. [5] define the occurrence of VIMS in simulators as simulator sickness and McCauley and Thomas [29] define CS as the occurrence of VIMS specifically in VR systems.

The CS, in turn, is comparable to the symptoms of MS occurring in the real world, such as nausea, vertigo, dizziness, stomach problems, and others [14]. Symptoms of CS mainly occur with HMD (Oculus Rift, HTC Vive, among others) [40].

Contrary to the previous works, in this work we are interested not only in predicting the CS moments but also in identifying the causes that are leading to such a discomfort. For that, we propose a novel classification solution for predicting the level of discomfort while using HMD during the gameplay. The literature still lacks a comprehensive description of CS and related topics, with most works consisting of empirical observations and reports [47], [26], [16].

As shown in Figure 1, we collected data on gameplay experiences when the user emerged out after finishing the VR game in addition to personal data captured through questionnaires. The subjective and objective data were then provided as input data to classifiers to predict the discomfort level of the user over the gameplay. The experiments were separated in binary and quarterly classifications. In the binary case, the labels specify whether there is discomfort or not. In the quarterly classification, we created four scales of discomfort situation: none, slight, moderate and severe discomfort. The performance of classifiers is analyzed and we also provide an attribute ranking analysis before presenting the conclusions.

## II. Related Work

In this section, we review related works, which are addressed in two categories related to CS: causes and prediction.

### A. *Causes*

Several factors can cause pain and discomfort when using HMD [48]. Manifestations of CS can lead to more intense symptoms, such as nausea, eye fatigue, neuralgia, and dizziness [18]. According to the literature [42], [27], [10], [23], it is possible to highlight the main factors that contribute to the manifestation of CS symptoms.

1) **Locomotion** - According to Rebenitsch, 2016 [40], locomotion speed may be correlated to CS. When the participant travels and if he has greater control of his movements and is close to natural movements then he will experience less CS.
2) **Acceleration** - Visual accelerations without generating any response in the corresponding vestibular organs cause uncomfortable sensations that result in CS symptoms. High accelerations during movements produce higher degrees of CS [25], [43].
3) **Field of View** - In VR environments, a wide field of view generates a great sense of immersion. However, a wide field of view contributes to the CS manifestation. In contrast, a narrow field of view creates a more comfortable experience in VR but decrease the user's immersion [48], [10].
4) **Depth of Field** - Inadequate simulation of focus on stereoscopic HMDs with flow tracking devices creates unbelievable images and, consequently, causes discomfort. In the human eye, focus forces blur effects naturally that depend on the depth of field (DoF) and distance range of objects in the observed area. Due to ocular convergence, objects outside this range, located behind or in front of the eyes, are blurred [35].
5) **Degree of Control** - According to Stanney and Keneddy [43], interactions and movements that are not being controlled by the user may cause CS.
6) **Exposure** - In a previous work, we showed that time exposure to VR experiences might raise discomfort in a proportional way [34].
7) **Latency—Lag**, has persisted for years as an obstacle in the previous generations of HMDs [31]. Latency is the delay between action and reaction latency is the time difference between the time of input given and the corresponding action to take place in a virtual scenario. High latency may drastically increase CS levels.
8) **Static Rest Frame** - The lack of a static frame of reference (static rest frame) can cause sensory conflicts and, ultimately, CS [6]. According to Cao et al. [6] most users are able to better tolerate virtual environments created by projectors such as cave automatic virtual environments (CAVEs) [8] compared to HMDs devices.
9) **Camera Rotation** - Rotations in virtual environments with HMDs increase the chances of sensory conflicts. The feeling of vection is greater in rotations when two axes are used in comparison to just one axis [4].
10) **Postural Instability** - Postural instability (Ataxia) is a postural imbalance or lack of coordination [25] caused when the body tries to maintain an incorrect posture due to the sensory conflict caused by the virtual environment. In other words, postural instability is the reactive response to information received by the vestibular and visual organs, which lead to CS.

### B. Cybersickness prediction

Some works [32], [21] have made use of subjective questionnaires to estimate and predict levels of discomfort. The most used questionnaires are SSQ, MSSQ and VRSQ.

Some works, [32] and [21], have made use of subjective questionnaires to estimate and predict levels of discomfort. The most used questionnaires are simulator sickness questionnaire (SSQ), MSSQ, and virtual reality sickness questionnaire (VRSQ).

Padmanaban et al. [32] designed a VR sickness predictor. In this approach, a dataset,is created with the some questionnaires to evaluate the physiological causes of sickness and individual historical elements to get a more precise result from users. They used the combination of two sickness questionnaires: MSSQ and SSQ, to find a single sickness value. They measured SSQ scores of various individuals through stereoscopic content. They used Flownet [9] to calculated optical flow vectors (they calculate optical flow from one frame to the next, which is measured pixel speed).

Kim, J. et al. [21], propose a deep learning architecture to estimating cognitive state using brain signals data and they are related to CS levels. They developed their model based on deep learning models (LSTM-RNN and CNN). The models learn the individual characteristics of the participants resulting in manifestation of CS symptoms when viewing a VR video.

Garcia-Agundez et al. [12] aimed to classify the level of CS. The proposed model used a combination of bio-signal and game settings. User signals, such as respiratory and skin conductivity of 66 participants were collected. As a result, they mentioned a classification accuracy of 82% (SVM) for binary classification and 56% (KNN) for ternary.

Jin et al. [17] separates factors that cause CS in three groups: hardware characteristics (VR device settings and features), software characteristics (the content of the VR scenes), and the individual user. The authors used classifiers to predict the level of discomfort. A total of three machine learning algorithms (CNN, LSTM-RNN, and SVR) were used. According to the results, the LSTM-RNN was the most viable model for the case.

The above-mentioned works do not classify the CS with actual data obtained during the gameplay. In Jin's work, the best result was achieved by recurrent neural networks. This is not a surprise, as the CS is linked to the amount of exposure time and also to a time series problem. Recurrent neural networks (RNNs) show good results for time series problems. All these works were focused on predicting the CS manifestation but not the causes. Different from other works, we used symbolic classifiers and it is possible now to analyze the discomfort patterns that give more details about the neural network decisions. In addition, in this work, we consider the entire VR experience, which means: before, during, and after the participation. We used data from questionnaires and VR interactions as well as the discomfort value entered by the user during the gameplay. Based upon these data CS level are symbolically classified with supervised learning.

## III. Cybersickness Prediction Approach

The manifestation of CS symptoms can occur due to several factors and configurations. As presented in the Related Works section, some of them use biological signals to quantify CS. Such data are already used by modern medicine to early detect diseases and malfunctions (i.e: heart problems). However, the use of specific equipment for capturing biological signals is intrusive and not practical.

As mentioned before, CS problems traditionally are quantified through discomfort questionnaires (such as SSQ, VRSQ) and susceptibility (by using the MSSQ). Such questionnaires are widely used nowadays and most of them focus their strategy for collecting subjective data of the state of the participant. However, in our research we noticed that the SSQ is not specifically aimed at detecting the CS but rather for quantifying the simulator sickness. For this reason, the use of VRSQ was chosen. However, no susceptibility identification questionnaires were used to quantify CS.

Moreover, for problems involving CS, the classifiers based on deep learning is proved to be a most suitable one, as CS problems associated with the time of using HMD devices are known. On the other hand, deep network classifiers are complex to understand. In other words, even if they produce a good final result, it is not trivial to discover the reasons for what reasons the neural network made such a decision.

In this work, we are interested in understand which are the most relevant attributes in observed CS situations and which are best symbolic classifiers. After this, the methodology of this research will be based on the combination of both subjective data (from users) and objectives information, collected during users' interaction with the VR environment. However, for a supervised machine-learning algorithm to learn, needs examples of outputs for a given set of attributes.

These output attributes are defined as classes. In the context of this work, the classes can be considered analogous to the four levels of SSQ discomfort (none, slight, moderate and severe). To capture data during the interaction and classify them in relation to the four levels of discomfort, we created interactive experiments where the user could interact and label in real time his level of discomfort. In addition to the data of the participants (subjective data), real time game data such as acceleration, head orientation, scene position among others, are also recorded. As far as we know, this combination of data is a totally novel approach in the context of data collection and machine-learning usage.

### A. Identification of relevant CS prediction data

Following the observations obtained through the experimentation stage, we proposed a preliminary dataset composed of 28 attributes obtained from the following sources: profile data, questionnaires data and gameplay data. All these features are captured taking into account two dependent variables: type of hardware and type of game. The complete subjective and objective parameters recorded can be observed in Table I.

The profile data was selected based on the literature and also on the experience acquired during pilot tests for this work.

TABLE I: Gameplay and users data.

| Feature set | | | |
|---|---|---|---|
| **Objective Data (Gameplay data)** | | **Subjective data (Users data)** | |
| **Feature** | **Type** | **Feature** | **Type** |
| Time Stamp | numerical | Gender | categorical |
| Speed | numerical | Age | categorical |
| Acceleration | numerical | VR Experience | categorical |
| Rotation (x, y and z) | numerical | Flicker Sensibility | categorical |
| Position (x, y and z) | numerical | Pre-symptoms | categorical |
| Region Of Interest | categorical | Glasses wearing | categorical |
| Size of FOV | numerical | Vision Impairments | categorical |
| Frame Rate | numerical | Posture | categorical |
| Static Frame | categorical | Dominant Eye | categorical |
| Haptic Feedback | categorical | Discomfort Level | numerical |
| Degree of Control | categorical | | |
| DoF Simulation | categorical | | |
| Player Locomotion | categorical | | |
| Automatic Camera | categorical | | |

We gathered this data through our CS Profile Questionnaire (CSPQ). The CSPQ contains nine questions such as:

- **Gender** - The gender of the participant is noted [2], [23] and women are more likely to experience visual discomfort compared to men.
- **Age** - We recorded the age of participants in three who are divided into three groups (18-36, 37-50, and above 50). According with studies, older participants are more susceptible to CS compared to younger ones [22], [39].
- **Experience in VR** - Level experience of the user with virtual environments was divided into two categories (without experience, with experience). According to Reason [37] sickness susceptibility is a product of the individual's overall experience with MS.
- **Flicker Sensibility** - Users are asked whether they feel discomfort when they are near the digital screens in order to discover the user's flicker sensibility. Flicker is a phenomenon of visual physiological discomfort and may cause physical and psychic fatigue in users in the vicinity of the disturbing load [41].
- **Pre-symptoms** - Health conditions can contribute to increased susceptibility to MS or CS when individuals are exposed to favorable environments. According to Frank et al. [11] and Laviola et al. [25], any symptoms, such as stomach pain, flu, stress, hangover, headache, visual fatigue, lack of sleep or respiratory diseases can lead to increased susceptibility to visual diseases.
- **Glasses Wearing** - According to Rebenitsch [39], vision correction can be correlated with CS and the correction of glasses for better view can create additional refraction of light and make a participant more able to feel CS symptoms.
- **Vision Impairments** - Although HMDs are compatible with the use of glasses or lenses by users with vision problems, we decided to ask if the user has one or more vision problems, such as myopia, hypermetropia or astigmatism.
- **Posture** - Postural instability is the reactive response to information incorrectly received by the vestibular and visual organs. According to Stroffegen et al. [44], the effects of postural instability precede CS symptoms, if they occur in VR environments. We also noted whether the user was seated or standing.
- **Dominant Eye** - According to Collins and Blackwell [7] most people have one dominant eye. In other words, the dominant eye sees more frequently and longer than the less dominant eye. HMDs show images to each eye simultaneously and separately. Because of this reason, we considered whether eye dominance has any connection to CS susceptibility and questioned the user for his dominant eye.

The questionnaire data contains information filled in by the user about discomfort symptoms before and after the experiment. The symptoms listed are from the VRSQ [20], which is a modified version of Kennedy's traditional SSQ to address specifically virtual reality environments with HMDs and the attributes considered are:

(1) The game data such as timestamp, speed, acceleration, player rotation axis, player position, the region of interest, size of the FOV, frame rate and discomfort level, class reported by the user at any time during the gameplay. (2) Boolean information such as existence of static resting frames, the existence of haptic response, level of user control over the camera, the existence of depth of field simulation (DoF) and whether the game primary camera moves automatically (without user intervention) or not.

### B. *Data classification*

In machine-learning, it is a common practice to validate the data and ensure that its format is valid before starting the training process. To ensure that, the collected and recorded raw data are converted into categorical values in a discretized pattern. Redundant attributes and objects in each data set are removed.

Our proposal is based on symbolic classifiers based on decision trees, such as random forest (RF). RFs or random decision classification, regression and other tasks that operate by constructing a multitude of decision trees at training time and outputting the class that is the mode of the classes or mean prediction of the individual trees. Using RF the input of each tree is tested from the original dataset. Moreover, a subset of features is randomly picked from these arbitrary features to improve the tree per node. Typically, random forest facilitates a wide number of inadequate classifiers to form a strong classifier [28].

We also evaluated other decision tree-based symbolic classifiers, such as BF Tree, CDT, Decision Strump, ForestPA, FT, Hoeffding, J48, J48 Graft, JCHAID Star, LAD Tree, Logistic model trees (LMT), Nb Tree, Random Tree, Rep Tree, and Simple Cart.

For validating the predictive model we use of k-fold cross-validation for all evaluated symbolic classifiers. Tripathi and Taneja [45] define k-fold cross-validation as a statistical method to estimate the machine-learning potential associated with a predictive model. This method is generally used for comparing and selecting a model for a given predictive modeling problem.

## IV. Experimental Data Capture

In order to collect experimental and real data, our proposal is based on capturing real information from real scenarios. For this work, two games 2: a race game and a flight game, were created. Both games will force the participant to perform habitual VR game movements, such as rotation, translation and perform acceleration changes. We captured data of the users in different places (school, university and technological events) for two months with two HMD devices (HTC Vive and Oculus Rift). All users agreed with their anonymous participation and assigned the consent terms.

For our experimental tests, each individual has to complete four tasks: filling the profile questionnaire (CSPQ), filling the VRSQ, participating in one of the VR games for up to 5 min (if possible) stating the numbers 0 (none), 1 (slight), 2 (moderate) or 3 (severe) each time when his/her level of discomfort is changed, and finally filling the VR sickness questionnaire VRSQ again. The participants were allowed to quit the experiment whenever they wanted.

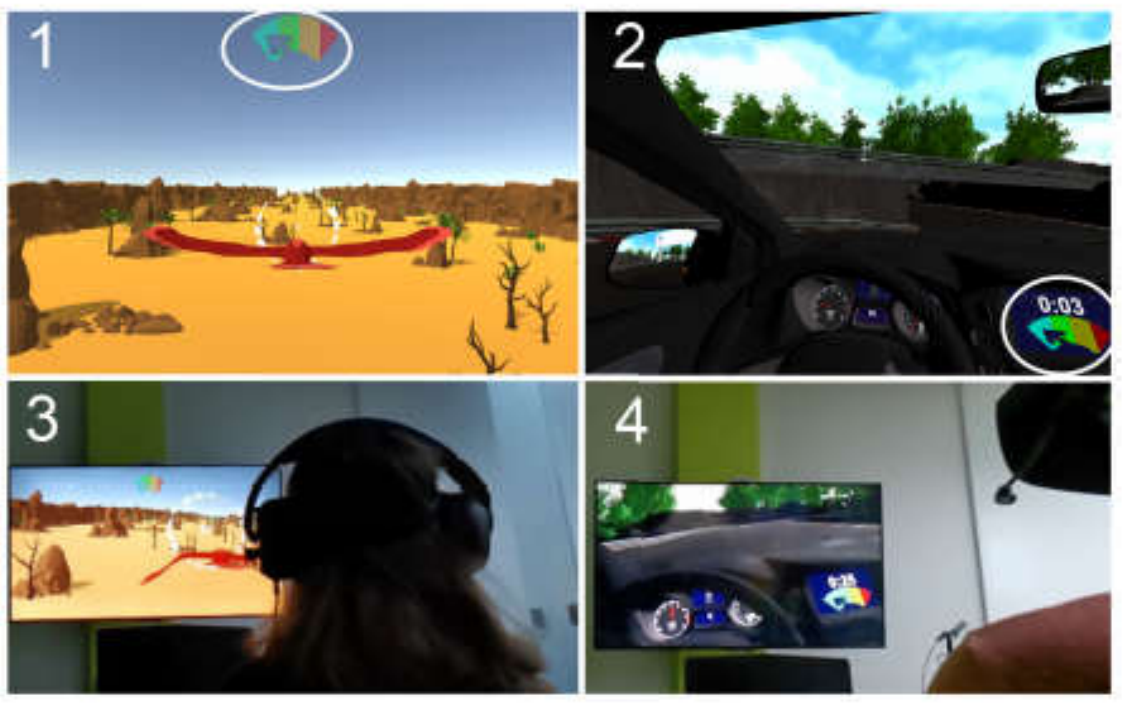

Fig. 2: Participants playing the flight game at the left (1 and 3) and the race game at the right (2 and 4). Participants are invited to state the numbers 0, 1, 2. or 3 according to their level of discomfort. In the Pictures 1 and 2 we can see a visual feedback (thermometer) of their current level of discomfort.

We validate data from 35 participants (nine female and 26 male) with ages ranging between 18 and 60 from a total of 86 participants achieved in the five phases of user tests. The complete participants profile can be seen in Table II. Although the experiment was carried out with a total of 86 participants considering all phases of testing, it is worth mentioning that only the valid data were considered, that is, data from users who answered all questionnaires correctly and completed the whole game interaction. Also the VRSQ data was used to validate inconsistencies between subjective and objective data captured. For instance, sometimes the user reported high levels of discomfort during the game interaction but in the post questionnaire pointed out none discomfort or vice-versa.

The definition of the data set to be captured has evolved over the stages. Based on what was raised in the literature, an initial set of attributes (possible causes) to be captured was defined, resulting both from the profiles of the participants and from the interaction with the games. However, this data set has been modified according to the experience acquired in the previous steps. Figure 3 shows the implemented functionalities, questionnaires and game data captured during the phases P1, P2, P3, P4 and P5.

TABLE II: Number of participants in each phase (P) of the experimental tests.

| Participants | P1 | P2 | P3 | P4 | P5 |
|---|---|---|---|---|---|
| **Gender** | | | | | |
| Male | 3 | 3 | 5 | 30 | 26 |
| Female | 1 | 1 | 1 | 7 | 9 |
| **Age** | | | | | |
| 18 to 36 | 2 | 4 | 6 | 33 | 28 |
| 37 to 50 | 2 | 0 | 0 | 4 | 4 |
| above 50 | 0 | 0 | 0 | 0 | 3 |
| **Hardware** | | | | | |
| Oculus Rift CV1 | 4 | 4 | 6 | 37 | 15 |
| HTV Vive | 0 | 0 | 0 | 0 | 20 |
| **Game** | | | | | |
| Race | 4 | 0 | 3 | 12 | 15 |
| Flight | 0 | 4 | 3 | 25 | 20 |

Recorded Attributes

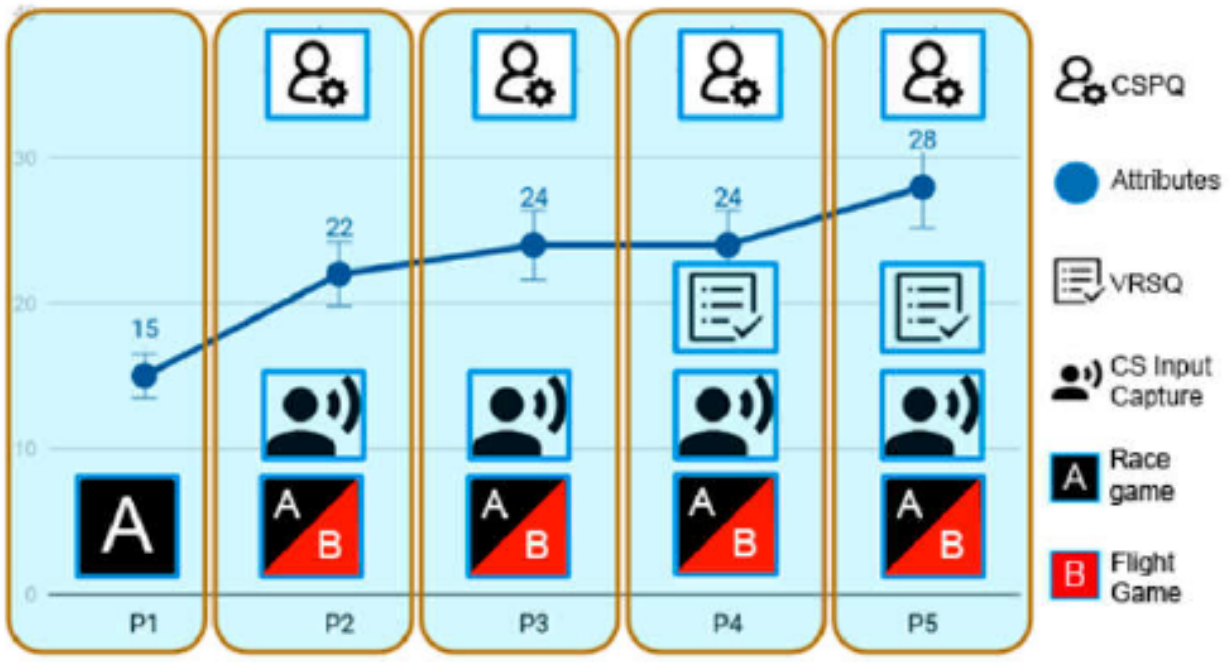

Fig. 3: This figure shows the five experimental phases and the available features in each of them. The complete list of the captured attributes can be seen in Table I.

## V. Classifiers

Once the data is collected, machine learning algorithms were trained to classify the CS. Each of the inputs stored by the experiment contains a rating given by users related to discomfort (from 0 to 3, where 0 is none and 3 is severe) during the gameplay.

To further analyze the collected data, we categorize the experiments into three main scenarios: A, B and C. These scenarios are:

- Scenario A classification consisting of data from the Racing game (3993 samples).
- Scenario B classification consisting of data from the Flight game (5397 samples).
- Scenario C classification using data from both scenarios together A and B (9390 samples).

Experiments were run using a 10-fold cross-validation in all scenarios. We also separated scenarios A, B, and C into two new groups. The first group is a binary classification (0-none or 1-discomfort, which includes from slight to severe classes). The second group is a quarterly classification containing all four classes (none, slight, moderate and severe). The distribution of classes can be seen in Figure 4.

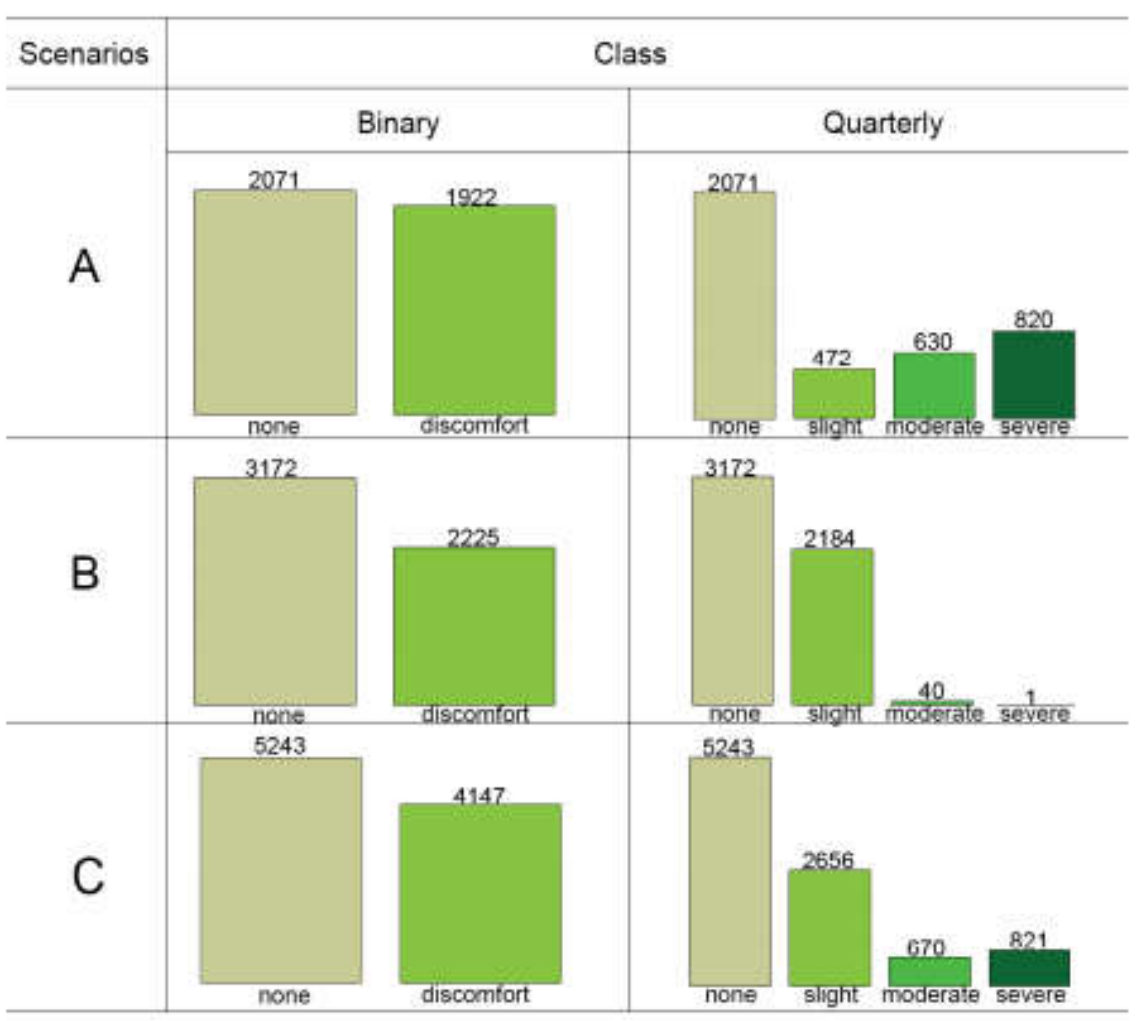

Fig. 4: Binary and quarterly class samples distribution in scenarios A (Racing game), B (Flight game) and, C (both games).

Tables III and IV show the accuracy and Kappa index for binary and quarterly classifications over scenarios A, B, and C.

## VI. RESULTS

This section presents the obtained results with the analysis of the captured data in the experimentation phases and usage of the developed immersive environments.

### A. *Binary classification*

As previously mentioned, we also merged the discomfort level into a single class in order to perform binary classifications, which are usually stronger than non-binary ones. Discomfort values that were previously represented as slight, moderate, and severe are represented as discomfort.

For all scenarios, that is, Scenario A, B, and C, the RF classifier is proved to be the best with an accuracy of 94.0%, 99.0%, and 96.6% for the binary case.

### B. *Quarterly classification*

For the quarterly classification, the dataset was not changed. This set of experiments were conducted based on four classes: none, slight, moderate and severe. In Scenario A, the classifier LMT achieved the best result, with accuracy of 92.4%. In the Scenario B, the best result was obtained with RF (98.9%). For Scenario C, RF also obtained the best classification accuracy (95.4%).

TABLE III: Binary Classification

| Binary Classification | | | | | | |
|---|---|---|---|---|---|---|
| Scenarios | A | | B | | C | |
| Classificator | ACC | KPP | ACC | KPP | ACC | KPP |
| BFTree | 91.8% | 0.8364 | 96.8% | 0.9353 | 93.0% | 0.8584 |
| CDT | 89.7% | 0.794 | 96.8% | 0.9346 | 92.3% | 0.8457 |
| DecisionStrump | 62.0% | 0.2574 | 71.9% | 0.4299 | 61.6% | 0.2713 |
| ForestPA | 91.5% | 0.8297 | 97.8% | 0.9563 | 95.5% | 0.909 |
| FT | 87.0% | 0.7395 | 94.2% | 0.881 | 90.9% | 0.8154 |
| Hoeffding | 69.9% | 0.393 | 78.1% | 0.5363 | 71.9% | 0.4251 |
| J48 | 92.4% | 0.8495 | 97.9% | 0.9576 | 95.2% | 0.9036 |
| J48Graft | 92.6% | 0.853 | 97.9% | 0.9575 | 95.2% | 0.9036 |
| JCHAIDStar | 89.8% | 0.7968 | 92.8% | 0.894 | 91.6% | 0.8582 |
| LADTree | 78.9% | 0.5812 | 88.9% | 0.7722 | 74.1% | 0.4829 |
| LMT | 93.0% | 0.86 | 98.1% | 0.961 | 95.5% | 0.9088 |
| NbTree | 88.1% | 0.7624 | 98.6% | 0.9728 | 95.0% | 0.8993 |
| RandomForest | 94.0% | 0.8805 | 99.0% | 0.9801 | 96.6% | 0.9323 |
| RandomTree | 89.2% | 0.7838 | 96.6% | 0.93 | 92.2% | 0.8421 |
| RepTree | 90.7% | 0.8147 | 96.9% | 0.9368 | 93.0% | 0.8595 |
| SimpleCart | 92.2% | 0.8455 | 97.2% | 0.9441 | 93.4% | 0.8672 |

TABLE IV: Quarterly Classification

| Quarterly Classification | | | | | | |
|---|---|---|---|---|---|---|
| Scenarios | A | | B | | C | |
| Classificator | ACC | KPP | ACC | KPP | ACC | KPP |
| BFTree | 88.8% | 0.827 | 97.1% | 0.9417 | 93.0% | 0.8821 |
| CDT | 86.3% | 0.7854 | 96.5% | 0.93 | 92.3% | 0.8698 |
| DecisionStrump | 51.0% | 0 | 71.6% | 0.4294 | 55.8% | 0 |
| ForestPA | 87.6% | 0.8017 | 97.4% | 0.947 | 94.2% | 0.9015 |
| FT | 83.1% | 0.7375 | 93.7% | 0.8731 | 89.4% | 0.8227 |
| Hoeffding | 52.5% | 0.0401 | 73.8% | 0.4882 | 55.8% | 0 |
| J48 | 90.7% | 0.8566 | 97.8% | 0.9569 | 94.8% | 0.9139 |
| J48Graft | 90.9% | 0.8598 | 97.7% | 0.9547 | 95.0% | 0.9162 |
| JCHAIDStar | 77.7% | 0.6513 | 92.3% | 0.8793 | 86.0% | 0.7824 |
| LADTree | 68.5% | 0.4996 | 87.0% | 0.7333 | 72.1% | 0.4694 |
| LMT | 92.4% | 0.8832 | 97.8% | 0.9566 | 95.5% | 0.9249 |
| NbTree | 88.7% | 0.8246 | 98.7% | 0.9747 | 94.4% | 0.9052 |
| RandomForest | 92.2% | 0.8782 | 98.9% | 0.9792 | 95.4% | 0.9221 |
| RandomTree | 85.1% | 0.7709 | 96.8% | 0.9355 | 89.5% | 0.8243 |
| RepTree | 87.0% | 0.7962 | 96.7% | 0.9328 | 92.6% | 0.8755 |
| SimpleCart | 88.9% | 0.8288 | 97.3% | 0.9464 | 93.0% | 0.8821 |

### C. *Attribute evaluation*

As we are interested in better understanding of the causes involved in the discomfort event, we have to evaluate the attributes involved in the CS prediction decision. For that, we used the Classifier Attribute Evaluator, after training it with the full set training and the "leave one attribute out" method of Weka [13]. This strategy generated a ranking of all attributes using the best classifier of the previous experiments (Tables III and IV). This algorithm removed attributes from the dataset and evaluates how its removal influences the performance of the classification. After all attributes were evaluated, they were ranked in terms of impact. The best ones stay at the top of the ranking.

For binary classifications and Scenario A (racing game), the most relevant attributes considered were (Table V) : time Stamp (time exposure amount), age, gender, rotation on the z-axis, and player speed. For Scenario B (flight game), attributes are as follows: age, VR experience, vision impairment, rotation on the z-axis, and time stamp. For Scenario C (both games), time stamp, age, gender, and player speed.

TABLE V: Binary attribute evaluation

| Binary Attribute Ranking (Leave One Attribute Out) | | | |
|---|---|---|---|
| Rank | Scenario A | Scenario B | Scenario C |
| 1 | Time Stamp | Age | Time Stamp |
| 2 | Age | Position Z | Age |
| 3 | Gender | VR Experience | Gender |
| 4 | Rotation Z | Vision Imp. | Player Speed |
| 5 | Player Speed | Rotation Z | Position Z |
| 6 | VR Experience | Time Stamp | VR Experience |
| 7 | Rotation X | Rotation X | Vision Imp. |
| 8 | Rotation Y | Speed | Rotation Z |
| 9 | Eye Dominance | Eye Dominance | Rotation X |
| 10 | Region of Interest | Position Y | Camera Auto |

TABLE VI: Quarterly attribute evaluation

| Quarterly Attribute Ranking (Leave One Attribute Out) | | | |
|---|---|---|---|
| Rank | Scenario A | Scenario B | Scenario C |
| 1 | Time Stamp | Age | Time Stamp |
| 2 | Age | Position Z | Age |
| 3 | Gender | VR Experience | Gender |
| 4 | VR Experience | Vision Imp. | Rotation Z |
| 5 | Speed | Time Stamp | VR Experience |
| 6 | Rotation Z | Rotation X | Player Speed |
| 7 | Eye Dominance | Rotation Z | Position Z |
| 8 | Rotation Y | Speed | Vision Imp. |
| 9 | Rotation X | Gender | Rotation X |
| 10 | Position X | Eye Dominance | Position Y |

These ranking of positions of attributes for the different scenarios are also validated by the literature [25], [4], [38], [2]. For two out of three test scenarios, the time stamp was considered the most important for the classification of discomfort. Player speed, rotation on z-axis as well as gender and age were also essential attributes [2].

In quarterly attribute evaluation (Table VI), we obtained very similar results. For Scenario A: time stamp, age, gender, VR experience, and player speed were the top five. In Scenario B, the most relevant attributes were: age, VR experience, and vision impairments. In Scenario C: time stamp, age, gender, rotation on the z-axis and VR experience are important attributes [2].

Considering the gender attribute analysis we observed from the captured data of the race game (3993 samples, from 15 participants where seven are females and eight are males) that female individuals reported lower incidents of discomfort compared to male participants, as shown in Figure 5. This finding disagrees with literature in which Biocca [2] and Kolasinski [23] report that female individuals are more susceptible to symptoms of MS. However, such behavior was only observed in MS scenario and not in CS scenario. Anyways these findings need to be further investigated taking into account more samples and also with other games. Besides that, we were able to observe this effect only in the race game because in the flight game the gender data was not well distributed.

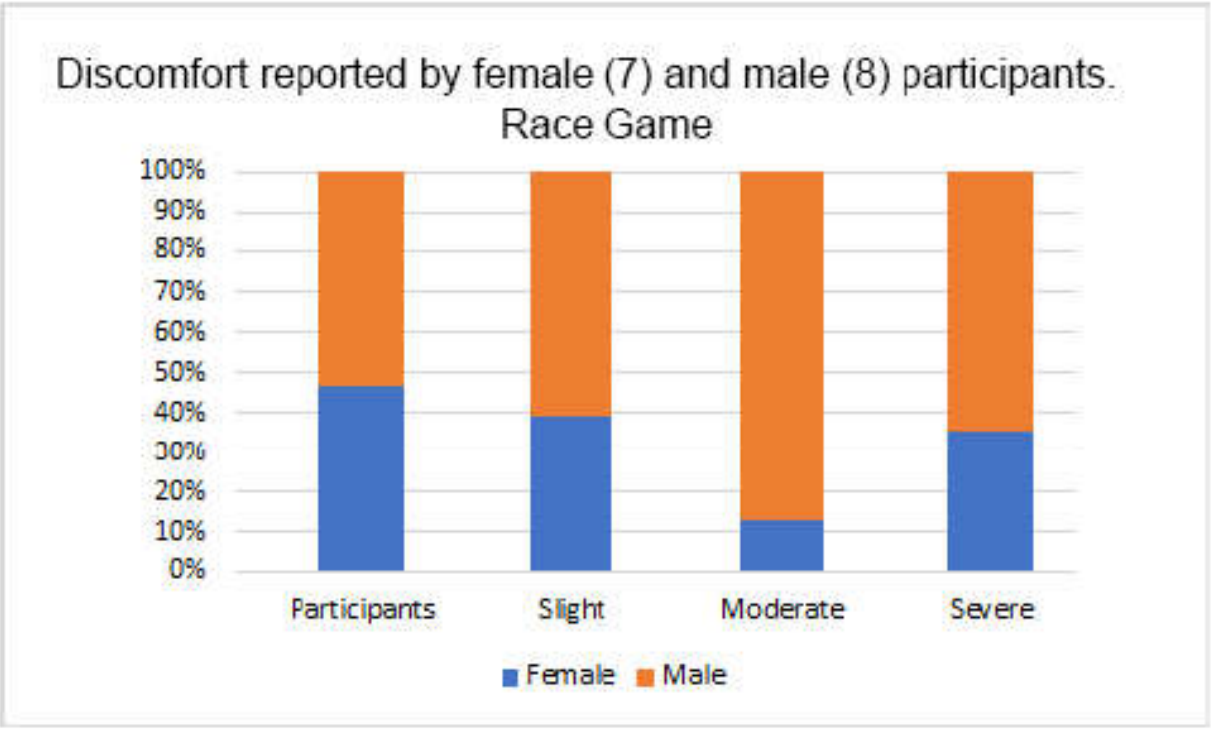

Fig. 5: Comparison of the discomfort levels reported by female and male participants in the race game.

## VII. Conclusion

As far as we know, this is the first work that uses classifiers in order to predict discomfort during the gameplay experience and evaluate a broad feature set, which includes user's personal data. We also share and provide the user data publicly [33].

Subsequently, the analysis in terms of machine learning consisted of three scenarios: Scenario A (data from the racing game), Scenario B (data from the flying game), and Scenario C (data from both games). We performed supervised binary and quarterly classifications using 16 decision tree classifiers. Classifiers that resulted in the highest accuracy were RF and LMT. The best accuracy was 99.0% and was obtained with the random forest classifier for Scenario B (flight game) in the binary classification.

An attribute selection was also performed in order to identify the most relevant attributes. For all scenarios it was observed that the most relevant attributes were the same, and they were exposure time, z-axis rotation and profile attributes of the individual (gender, age, and VR experience). These results corroborate the importance of attributes related to the individuals in the prediction of CS. This assessment reinforces the theories and hypothesis present in the literature known so far [34], [25], [1], [22]. Duration of the experience (time) is notably essential for the prediction of discomfort in VR environments.

Future works may increase the number of individuals in the experimental trials and also include new attributes. Due to the limitation of the number of games developed (only 2), some subjective and objective features such as the player's posture, locomotion, and others were not included in the training. To overcome the above issues, we also intend to produce other types of games and conduct new user experiment tests in future.